\RequirePackage{fix-cm}
\RequirePackage{booktabs}
\RequirePackage{amsmath,amssymb,amsfonts}
\RequirePackage[ruled,vlined,linesnumbered]{algorithm2e}

\documentclass[smallextended]{svjour3}       
\smartqed  
\usepackage{graphicx}
\usepackage{xcolor}
\begin{document}

\title{PASSVM: A Highly Accurate Online Fast Flux Detection System
}


\author{Basheer Al-Duwairi* \and Moath Jarrah \and Ahmed Shatnawi         
}


\institute{Basheer Al-Duwairi \at
              \email{basheer@just.edu.jo}           
           \and
           Moath Jarrah \at
              \email{mjarrah@just.edu.jo}              
           \and
           Ahmed Shatnawi \at
           \email{ahmedshatnawi@just.edu.jo}\\ \\
           Jordan University of Science \& Technology\\
           Irbid 22110, Jordan \\
           Tel.: +962-02-7201000\\
           Fax: +962-02-7201077
}

\date{Received: \today}

\maketitle

\begin{abstract}
Fast Flux service networks (FFSNs) are used by adversaries to achieve a high resilient technique for their malicious servers while keeping them hidden from direct access. In this technique, a large number of botnet machines, that are known as flux agents, work as proxies to relay the traffic between end users and a malicious mothership server which is controlled by an adversary. Various mechanisms have been proposed for detecting FFSNs. Such mechanisms depend on collecting a large amount of DNS traffic traces and require a considerable amount of time to identify fast flux domains. In this paper, we propose an efficient AI-based online fast flux detection system that performs highly accurate and extremely fast detection of fast flux domains. The proposed system, called PASSVM, is based on features that are associated with DNS response messages of a given domain name. The approach relies on features that are stored in two local databases, in addition to features that are extracted from the response DNS messages itself. The information in the databases are obtained from Censys search engine and IP Geolocation service. PASSVM is evaluated using three types of artificial neural networks which are: Multilayer Perceptron (MLP), Radial  Basis  Function  Network  (RBF), and Support  Vector Machines (SVM). Results show that SVM with RBF kernel outperformed the other two methods with an accuracy of 99.557\% and a detection time of less than 18 ms. 


\end{abstract}


\maketitle

\section{Introduction}

The Domain Name System (DNS) is a core Internet infrastructure element that is implemented as a distributed hierarchical database and is viewed as a crucial backbone of the Internet. It mainly provides a mapping between domain names and their IP addresses, in addition to other important functions that are necessary for the proper functions of websites. At the same time, DNS plays an integral part in the operation of different types of attacks and malicious activities such as DNS amplification attacks \cite{macfarland2017best} \cite{anagnostopoulos2013dns}, DNS cache poisoning attacks \cite{alharbi2019collaborative} \cite{jackson2009protecting}, malware distribution \cite{zhauniarovich2018survey}, and botnets \cite{singh2019issues} \cite{patsakis2020encrypted}. As millions of new domain names are registered every day, there is a growing concern that many of these domains might belong to botnets and various types of malicious activities. 

Botnets represent a significant threat that is continuously evolving with new techniques and architectures. They are used to perform different types of malicious activities such as distributed denial of service attacks, email spam, phishing, and malware distribution. Botherders are continuously developing techniques to hide their malicious activities and to evade detection. Attackers rely on DNS to resolve IP addresses of domain names (e.g., phishing domains, command and control (C\&C) servers, etc.) that are used in their attacks. Therefore, it is clear that DNS provides an important information that can reveal different attack activities. 

Fast Flux Service Networks (FFSNs) are spacial forms of botnets that are mainly designed to provide resilient and highly available service while evading detection. It is a technique that is adopted by botmasters since 2007 with an increasing activity rate in recent years. The main purpose of fast flux networks is to hide the content server (also called the mothership server), where the malicious content is hosted behind a botnet of compromised machines that are called flux agents. Flux agents are configured by the botmaster in order to serve as proxies that relay traffic to/from the origin server. Similar to content distribution networks (CDNs), FFSNs achieve a high availability by using a technique that mimics the Round Robin Domain Name System (RRDNS) to map domain names and IP addresses of flux agents. In FFSNs, a fast flux domain is mapped to multiple IP addresses that keep changing very fast. This increases the chances that the origin server is reachable from some of the flux agents that are still running and have not blacklisted yet. 
 
The Internet Honeynet project was the first to systemically describe the problem of fast flux networks and their main features \cite{saluskyknow}. Subsequently, the research community paid more attention to this growing threat and several mechanisms were proposed to address the problem. With the renewed adoption of fast flux networks in major botnets (e.g., SandiFlux \cite{Sandiflux} and DarkCloud \cite{darkcloud}), new fast flux detection mechanisms were proposed in recent years (e.g., \cite{fastflux-passive-2} \cite{fastflux-passive-5}). Most of these mechanisms rely on analyzing DNS traffic information that corresponds to fast flux domains in order to characterize their behavior and identify their distinguishing features. In this regard, DNS records can be obtained actively by issuing DNS requests about domain names of fast flux domains that are obtained from email spam campaigns and phishing archives, or through the analysis of passively collected DNS traffic traces. 

Detecting fast flux networks accurately and instantly (i.e., online detection) is an important and a challenging problem. While some previously proposed mechanisms have achieved high detection rate, they require long time to collect DNS and other related information from different sources. In this paper, we propose a novel and efficient AI-based online fast flux detection system. The main goal of the proposed system (PASSVM) is to perform online fast flux detection based on a single DNS response message for a given domain name. To achieve this goal, we investigated different Artificial Neural Networks (ANN) models and trained them using features that are stored locally. The features information are based on the A records that are found in the DNS response message. This means that we restrict our feature set to the information that is available in the DNS response itself or can be obtained from local databases that can be downloaded in advance. Therefore, well-known fast flux features that require any form of active query are not needed. The first database is constructed from data that is downloaded in advance from Censys search engine and includes historical information about IPv4 address space. The second database is constructed in advance using data from IP geolocation service \cite{iptoasn} which provides IP to location and autonomous system number (ASN) information for IP addresses that correspond to a given domain name. Specifically, the main contributions of this paper are:
\begin{enumerate}

	\item An efficient and highly accurate online fast flux detection system based on features that are available in a single DNS response message.
	\item Leveraging the data that is made available by Censys search engine \cite{censysio} about IPv4  address space and data from IP geolocation service.
	\item Two new fast flux features that allow for online fast flux detection are proposed. The two features are based on the database that can be downloaded from Censys search engine.
	
	\item The proposed system is evaluated using three types of artificial neural network models which are: Multilayer Perceptron (MLP), Radial  Basis  Function  Network  (RBF),  and  Support  Vector Machines (SVM). SVM with RBF kernel outperformed the other two methods with an accuracy of 99.557\% and a detection time of less than 18 ms. 
	
\end{enumerate}

The remainder of the paper is organized as follows. Section \ref{sec:background} presents a background about fast flux networks. In section \ref{sec:previous}, we discuss the related work. In section \ref{sec:proposed}, we present the proposed PASSVM system. In Section \ref{sec:classification}, we discuss different ANN models that are used in the evaluation of the proposed approach. In Section \ref{sec:evaluation}, we evaluate our system and discuss the performance of different ANN models based on recent datasets. The conclusion of the paper is given in section \ref{sec:conclusion}.

\section{Fast Flux Networks\label{sec:background}}

Round robin DNS and content distribution networks (CDNs) are two main techniques that are employed by web servers to achieve load balancing and high availability. In round robin DNS, the authoritative domain name server of a certain domain name is configured to distribute the workload to multiple redundant web servers by mapping the host name of the web server to multiple IP addresses. This mapping keeps changing in a round robin fashion. Each time a client issues a DNS query, the client may obtain a list of IP addresses for the given host name in different order. In CDNs, the content is pushed to a large number of geographically distributed servers. Global load balancing is achieved by providing the client with set of IP addresses of nearby servers. For example, a user in USA, who is trying to access a CDN hosted website, sends a DNS query for that web site, and will get a reply with a set of IP addresses of servers that are hosted in nearby locations within the CDN.   

\begin{figure}
	\centering
	\includegraphics[scale=0.350]{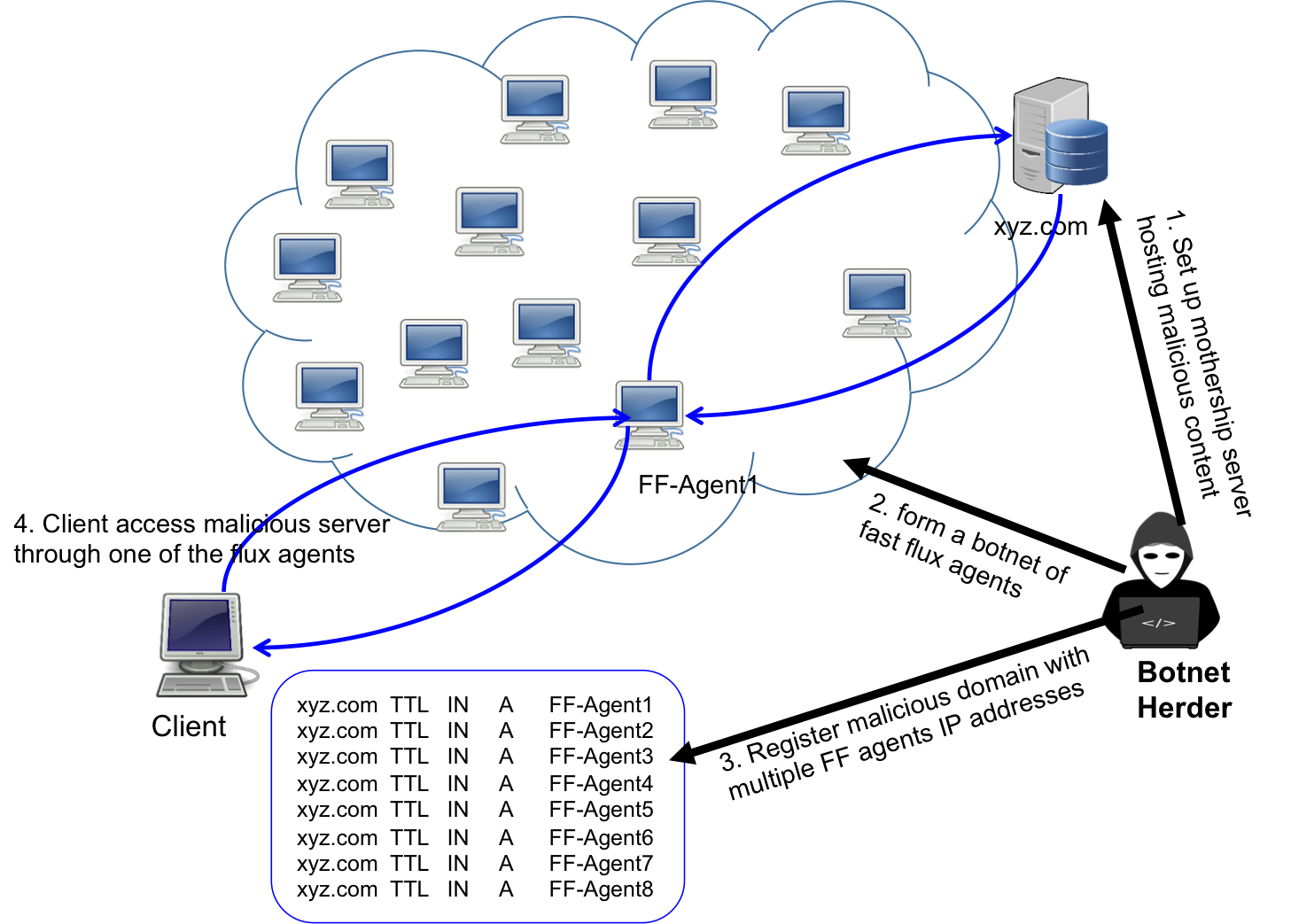}
	\caption{Stages of the life cycle of fast flux service network}
	\label{fastflux-operation}
\end{figure}

FFSNs employ similar techniques in order to provide a high availability of malicious servers while hiding their real locations. Figure \ref{fastflux-operation} shows the main stages of constructing and operating a fast flux service network. Initially, the botnet herder sets up a mothership server in order to host some sort of malicious content for the purpose of malware distribution, illegal pharmaceutical products sale, or hosting adult content, etc. (step 1). A domain name, such as \textit{xyz.com}, is assigned for this server. Then, a botnet of fast flux agents is formed and each agent is configured to serve as a proxy server to relay traffic to/from the mothership server (step 2). Flux agents are mainly compromised machines with intermittent connectivity, limited computational power, and low to average bandwidth.

Afterwards, the botnet herder registers the domain name of the FFSN with a set of IP addresses that belong to the fast flux agents botnet (step 3). Therefore, any access to the malicious FFSN domain should go through one of the flux agents that is returned to the client by the DNS system (step 4). It is clear that the botnet of flux agents forms a protection layer for the hidden malicious server. In order to increase the resilience of the network and to evade detection, the botnet herder keeps changing the domain name registration in a fast manner. This type of FFSNs is called a \textit{single-flux}. There is a more sophisticated type, that is called \textit{double-flux}, in which the botnet herder also changes the mapping between the authoritative name server of the FFSN and its IP addresses in a fast manner. Therefore, providing a layer of protection for the FFSN' authoritative name server.

\begin{figure}
	\centering
	\includegraphics[scale=0.50]{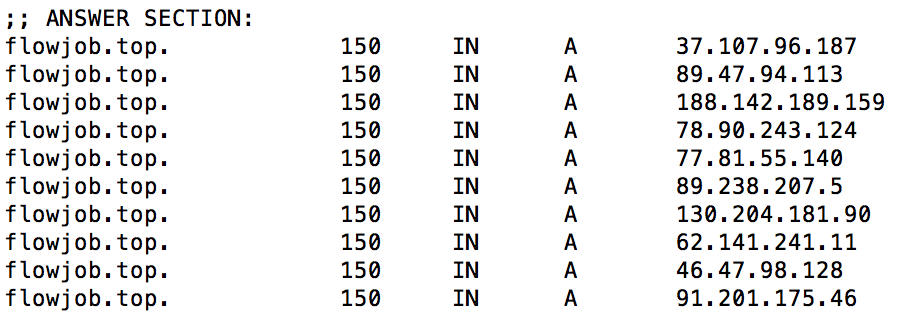}
	\caption{Output of the first dig of the fast flux domain \textit{flowjob.top.} (performed on April 22 2019)}
	\label{fastflux-dig1}
\end{figure}

\begin{figure}
	\centering
	\includegraphics[scale=0.50]{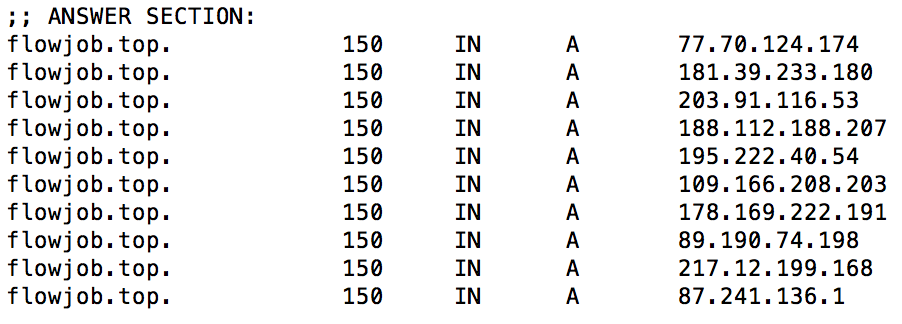}
	\caption{Output of the second dig of the fast flux domain \textit{flowjob.top.} (performed on April 22 2019)}
	\label{fastflux-dig2}
\end{figure}

Figures \ref{fastflux-dig1} and \ref{fastflux-dig2} show the DNS lookup result that is obtained using the Unix dig utility for the fast flux domain \textit{(flowjob.top.)}. It can be seen that this fast flux domain is mapped to multiple IP addresses and the mapping keeps on changing over time. For example, the second dig, which was performed $150$ seconds after the first one, showed a new set of IP addresses that did not appear in the first dig output, which is a common characteristic of fast flux networks. Previous research studies have identified several features that mainly characterize fast flux domains \cite{fastflux-active2} \cite{fastflux-passive-2}. These features include:

\begin{itemize}
	\item \textit{Large number of IP addresses.} The number of A records that is included within a single DNS response message of a fast flux domain is relatively large. If one or more of the fast flux agents that are associated with the IP addresses are down, a client, that is trying to access the mothership server of the associated domain name, would automatically try another IP address (i.e., another agent) until it succeeds. Registering the domain name with a large number of IP addresses provides high availability of the malicious server as it increases the probability that one of the flux agents is up and running. 
	\item \textit{Large IP growth.} In order to avoid blacklisting, the mapping, between a fast flux domain and agent IP addresses, keeps on changing over time. Therefore, the number of IP addresses, that are associated with a certain fast flux domain, becomes large. 
	\item \textit{Low TTL value.} Since the mapping between a domain name and IP addresses changes very fast in FFSNs, then the TTL values are kept low. This guarantees that the values expire soon after the fast flux domain is resolved in order for users to obtain the new list of IP addresses.
	\item \textit{Large number of autonomous systems.} The IP addresses, that are returned in response to a DNS query for a fast flux domain, represent compromised machines that belong to different organizations and Internet Service Providers. Therefore, it is expected that IP addresses of these agents belong to multiple autonomous systems. 
	\item \textit{Large number of countries.} Previous studies showed that the IP addresses of fast flux domains are usually located in relatively large number of countries. This is expected since attackers register their fast flux domains with a set of IP addressees that are selected randomly from a pool of fast flux agents. 
	\item \textit{Domain names do not last for a long time.} The life time of a fast flux domain is relatively short. Attackers tend to register a large number of domains for their FFSNs, where each domain name remains active for a short period of time.
\end{itemize}

\section{Related Work \label{sec:previous}}
Most of the previous work in the area of fast flux detection (e.g., \cite{chen2019deep} \cite{kokkelkoren2019catching} \cite{al2015gflux} \cite{wang2017hiding}) have focused mainly on analyzing DNS traffic traces. Some methods performed active DNS probing to collect DNS records about suspect domains; while other methods used DNS records that were collected passively. Active and passive DNS information collection are usually combined with other information that is collected from different sources such as whois database, IP2location services, and blacklisted domains. In addition, some information are based on active measurements of the delay and other parameters. 

Characterization of fast flux networks was first presented in \cite{saluskyknow} and \cite{fastflux-active-1}. In these studies, active DNS-based approach was proposed where the DNS system was queried actively for domain names that were collected from Internet Spam archives and obtained by means of spam traps. DNS A records were analyzed by searching for fast flux domain footprints. These studies provided important insight about the nature of this threat and identified the main characteristics of fast flux domains such as such as low TTL values, large number of IP addresses, geographical distribution of flux agents, sharing of flux agents, and sharing of scam web pages. In \cite{fastflux-active-1}, a metric, which is called fluxy-score, is defined and computed over a set of parameters that are related to DNS records for a certain domain in order to determine whether the domain is a fast flux domain or a legitimate. 

Konte, Feamster, and Jung revealed similar characteristics and provided an insight about the dynamics of scam hosting infrastructure with a focus on the role of fast flux service networks \cite{fastflux-active2}. Other studies (e.g., \cite{fastflux-active3} \cite{fastflux-active4}) have focused on botnet detection through fast flux identification. The main problem of active DNS-based detection of fast flux networks is that it incurs high delay because it requires long time to collect DNS records for suspicious domain names. Also, the high rate of DNS queries that is received by authoritative domain name servers, that are under attackers' direct control, may alert them about such activity.  
 
Other fast flux detection mechanisms (e.g., \cite{FluxBuster} \cite{fastflux-passive-2}) relied on passive DNS monitoring, where DNS A records are obtained in a passive manner through monitoring DNS traffic. In \cite{FluxBuster}, live DNS traffic traces were captured by placing sensors at various strategic locations within an ISP network. The c4.5 decision tree classifier was applied on a set of features that are extracted from the traces and other features that are obtained actively. The reported results showed a high false positive rate and a long detection time due to the requirement of monitoring domains for a long period of time (five days in some cases). The work presented in \cite{fastflux-passive-2} followed a similar approach where DNS traffic traces were collected from a large corporate network. Mathematical and data mining techniques were applied on a set of features that are extracted from the monitored traffic in order to achieve near real-time fast flux detection. A system called Fast-flux hunter was proposed in \cite{fastflux-passive-5}. The system combines supervised and unsupervised online knowledge learning based system for fast flux detection based on features extracted from passive DNS traffic. In \cite{fastflux-passive-4}, a hybrid fast flux detection method that combines real-time detection and ong term DNS traffic analysis and monitoring was proposed. The method employed a decision tree classifier to achieve an accuracy rate close to 96\%.

Generally, a passive fast flux detection approach does not involve direct interaction with the domain name system. This has the advantage of eliminating network delays in obtaining DNS records. Also it prevents false DNS replies that can be provided by attackers who might be controlling authoritative domain name servers while observing a large number of DNS quires. Moreover, it has the advantage of discovering fast flux domains that could potentially appear in different malicious sources such as phishing emails, hackers forums, and online social networks. However, this approach requires processing a huge amount of DNS traffic traces that contain different types of information regarding malicious domains and legitimate domains.

The approach presented in  \cite{fastflux-realtime1} does not rely mainly on collecting DNS information. Instead, it relies on certain intrinsic characteristics of fast flux networks with the observation that it is expected to have long delays when fetching a document through a flux agent. The flux agent acts as a proxy to a back-end mothership server. On the other hand, the required time to download a similar content from a legitimate server is short. The scheme can detect fast flux domains within few seconds whenever a client attempts to download a document from a certain server. The scheme involves issuing additional HTTP requests to verify the legitimacy of the web-server. Also it can be applied in an active mode to identify fast flux domains in advance. However, this scheme has several limitations as it requires live interaction with malicious servers, and it does not discover flux agents that are hosted on a powerful PCs. Also it can result in many false positives in case legitimate servers are hosted at low-level hardware.

Previous fast flux detection mechanisms suffer from major drawbacks in the sense that they usually require considerable amount of time to actively or passively collect information about the flux DNS. Also, the mechanisms can not detect new fast flux domains before collecting enough data about them. Therefore, the DNS traffic analysis requires a long time of computations in order to achieve acceptable detection accuracy. Different than the previous work, the proposed fast flux detection system of this paper performs an on-the-fly highly accurate fast flux domain detection by leveraging information about IPv4 address space that is obtained in advance and stored in local databases. As a result, it eliminates the long-time monitoring and analysis of the DNS traffic and makes it possible to detect fast flux domains using only a single DNS response message.

\section{The Proposed PASSVM System \label{sec:proposed}}
	
\subsection{Overview of the proposed system}

An online fast flux detection system should be able to perform fast flux detection on the fly based on the available A records in a single DNS response message for the domain name. In other words, the system should directly answer whether a domain name is a fast flux domain or not without performing active DNS probing or seeking additional information from external sources during the decision making time. This requirement is very important to avoid delays and prevent additional traffic overhead during the classification process. To achieve this goal, PASSVM relies mainly on information about IPv4 addresses that are collected in advance (for example, one day before performing fast flux classification) from two main sources. The first source is the Censys search engine \cite{censysio}, which is a search engine that performs Internet wide scanning using the Zmap fast Internet scanner \cite{censys15}. Censys performs a daily IPv4 address space scanning and can be obtained by users using special APIs. Alternatively, Zmap can be used directly to perform the daily scanning in advance since it has the ability to scan the whole IPv4 address space within less than one hour \cite{zmap}. The second source is the IP geolocation service \cite{iptoasn}, which provides the mapping between IP addresses and their locations (cities and countries). It also provides the ASN number for each IP address. IP geolocation data can be downloaded and used locally. In addition, the proposed approach uses specific features that are extracted from the DNS response message of the request. 

\begin{figure}
	\centering
	\includegraphics[scale=0.40]{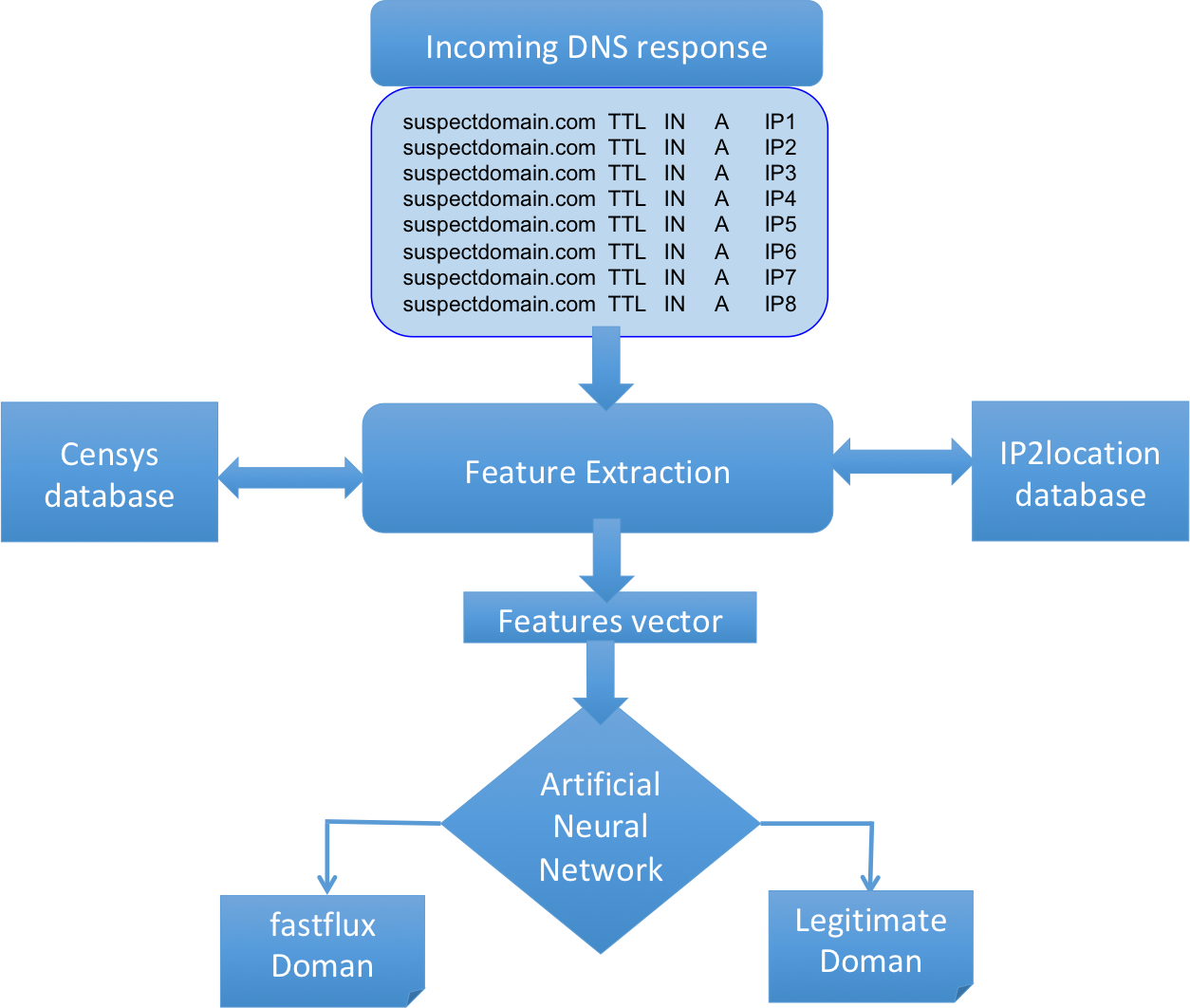}
	\caption{The proposed PASSVM system architecture}
	\label{system}
\end{figure}

Figure \ref{system} shows the proposed system architecture. The system can be used as a module within a local DNS resolver, where suspicious DNS requests are inspected by the PASSVM. Any domain name, that has many IP addresses (for example , more than 5 IPs), is considered a suspicious domain. As depicted in Figure \ref{system}, a feature set is extracted from the DNS response and from the local databases that were constructed using Censys and IP geolocation. An artificial neural network classifier is then used to decide whether the domain is a fast-flux domain or a legitimate domain.  

\subsection{Fast flux features set}

In this subsection, we describe the fast flux features that are used in the detection process of the proposed AI-based system. In order to achieve a fast classification model, we used features that can be obtained from the locally stored IPv4 address space information or from the DNS response message. 

\subsubsection{Censys-based fast flux features} 
Two important fast flux features that complement other known fast flux features are introduced in this paper. Here we discuss the two new features. Given a DNS reply of a fast flux domain name, it is expected that some of the flux agents, of the list of IP addresses in the DNS reply, might be down at a given time. The reason is that these agents are machines that belong to normal end users in different organizations and can be powered off or get disconnected from the Internet at any time. Consequently, querying Censys with a set of IP addresses that belong to a fast-flux domain will not return information about all the addresses in the search query. This is due to the fact that Censys data is obtained by performing a daily Internet wide scanning for the IPv4 address space using Zmap fast Internet scanner, and there is a high probability that the scanner will not find some of the fast flux agents because they are offline at the time of the scanning. On the other hand, IP addresses that correspond to a legitimate domain name are usually well-maintained servers. Therefore, they are not expected to be offline and there is a high probability that they are reachable by the Zmap scanner. This means that querying Censys with a list of IP addresses that correspond to a legitimate domain will return information about most of them. Hence, the ratio of IP addresses that are returned from Censys to the number of IP addresses that are submitted in the query represents an important feature to distinguish between fast flux domains and legitimate domains. This is the first new feature that is introduced in this paper.

This second important feature, that can be extracted from Censys search results, is related to the overall number of open ports that are discovered by Censys for the set of IP addresses of a certain domain name. In the case of a legitimate domain name, it is expected that all hosting servers, of the same domain, have similar configurations that result in having the same open ports on the hosting servers. On the other hand, in the case of a fast flux domain, there is a high chance that other port numbers are open, in addition to the ones that are configured by the attacker. This is because the configuration of the infected machines in the FFSNs are heterogeneous and belong to many different users. Hence, the number of open port numbers can indicate whether a domain name is a fast flux or a legitimate domain.   

\begin{figure}
		\centering
		\includegraphics[width=\columnwidth]{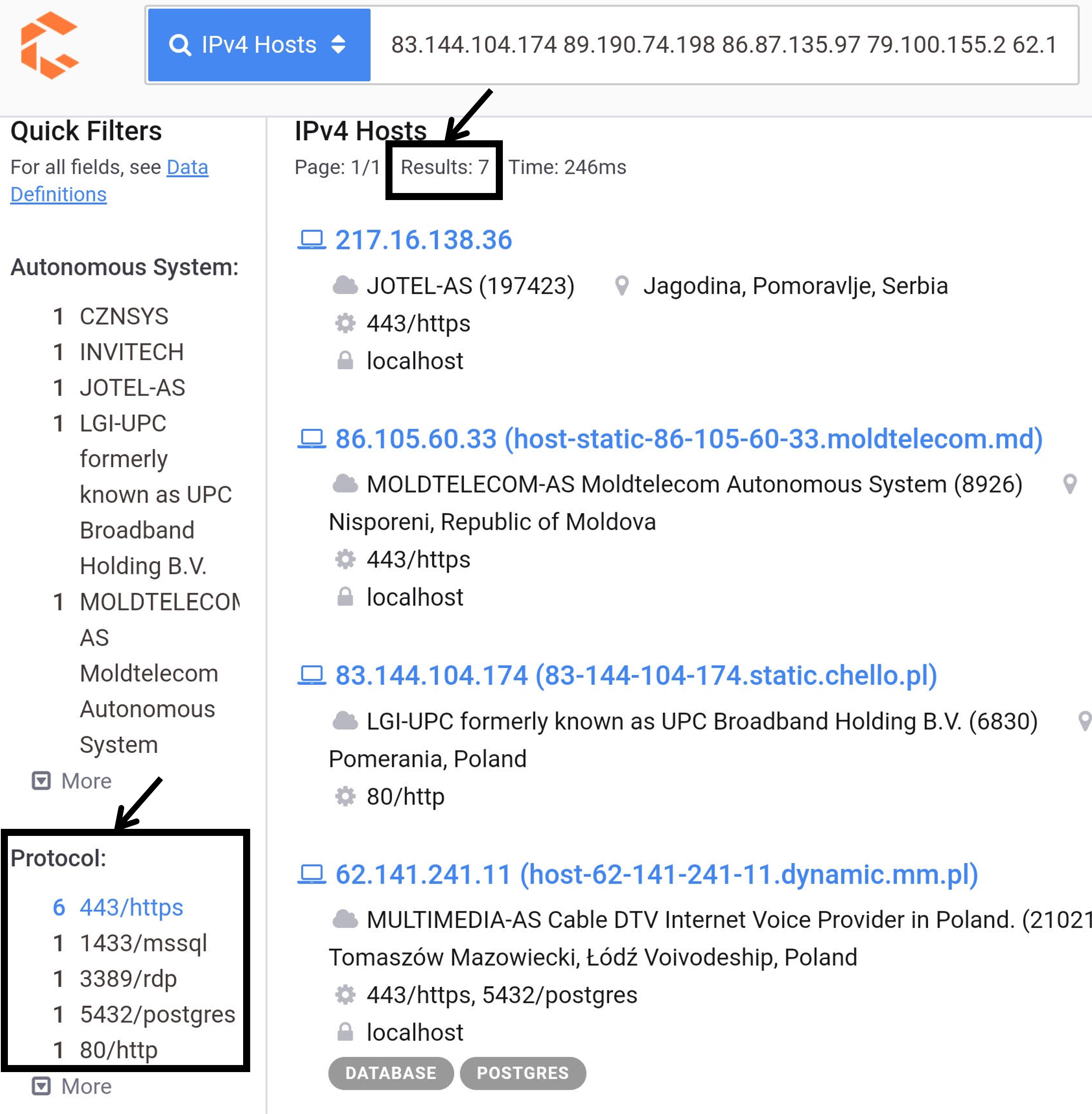}
		\caption{Example of the results that are returned by Censys for 10 IP addresses of the fast flux domain \textit{hex001.info.} (performed on December 7 2019)}
		\label{censys-fastflux}
\end{figure}

\begin{figure}
		\centering
		\includegraphics[width=\columnwidth ]{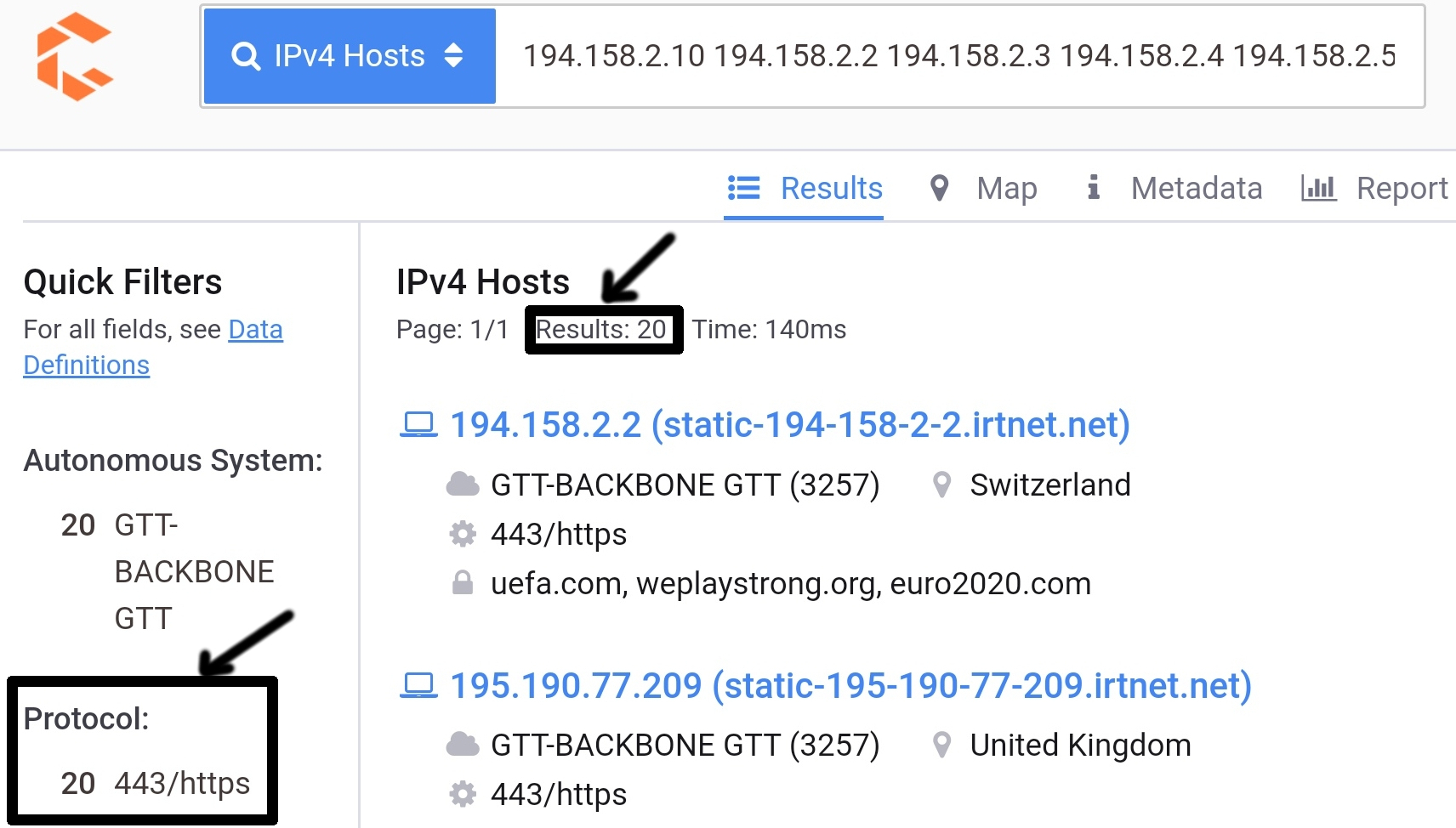}
		\caption{Example of the results that are returned by Censys for 20 IP addresses of the legitimate domain \textit{uefa.com.} (performed on December 7 2019)}
		\label{censys-legitimate}
\end{figure}

For illustration, Figure \ref{censys-fastflux} shows the results that are returned from the Censys search engine for a set of IP addresses that belong to the fast-flux domain (\texttt{hex001.info.}). Censys has returned information about $7$ IP addresses out of the $10$ IP addresses that were submitted in the query. On the other hand, Figure \ref{censys-legitimate} shows the results that are returned from the Censys search engine for a set of IP addresses that belong to the legitimate domain (\texttt{uefa.com.}). As shown in the figure, Censys has returned information about all of the IP addresses that were submitted in the query. Hence, the IP ratio of the fast-flux domain (\texttt{hex001.info.}) is $0.7$, and the IP ratio of the legitimate domain (\texttt{uefa.com.}) is $1.0$. In addition, the figures show the number of open ports of the IP addresses in the query. For the fast flux domain, there are five distinct open port numbers (Ports 443, 3389, 1433, 5432, and 80). However, for the legitimate domain there is only one open port number (Port 443). Hence, the number of open ports of fast flux domains is relatively greater than that of legitimate domains. 
    
In summary, the two newly introduced features for fast-flux detection are: 
    \begin{itemize}
        \item \textit{IP ratio:} The ratio of the number of IP addresses that is returned from Censys to the number of IP addresses that is submitted in the query.
        \item \textit{Ports:} The number of distinct open port protocols for all of the IP addresses that are returned from Censys search engine. 
    \end{itemize}

\subsubsection{IP geolocation-based features} Censys search engine provides information about geographical distribution of the IP addresses in a query to its database. This includes information such as countries, cites, and ASN numbers. However, sometimes Censys does not necessarily provide this information about all of the IP addresses that are found in a given DNS response. Therefore, IP geolocation service is used to obtain the information for all of IP addresses in a DNS response message that belongs to a given domain name. In particular, we define the following three features that are based on the number of countries and the number of ASNs provided by the IP geolocation service:
\begin{enumerate}
\item \textit{$ASN_{ratio}$:} This feature is defined as the ratio of the number of distinct ASNs for the set of IP addresses in a given DNS response message to the total number of IP addresses in the DNS response. For example, if the number of IP addresses that is returned in a DNS response message for a certain domain is $10$, and the IPs belong to $5$ distinct ASNs, then the $ASN_{ratio}$ equals to $0.5$. 
\item \textit{Regions:} This feature defines the number of distinct countries for all of IP addresses in a given DNS response message.
\item {RegionalSpread:} This feature is defined as the ratio of the number of distinct countries, for all of IP addresses in a given DNS response message, to the number of IP addresses in the response message.  
\end{enumerate}

\subsubsection{DNS-Response based features} The DNS response message itself contains important features that contribute significantly in distinguishing fast-flux domains. This includes the following features:
\begin{enumerate}
\item \textit{DomainLeangth:} This feature is defined as the number of characters in the domain name. Usually, malicious domains, including fast flux domains, have long domain names. Therefore, the domain name length is included as one of the features for fast flux detection.  
\item \textit{IPCount:} This feature is defined as the number of A records that are found in a DNS response message for a given domain. As explained in Section \ref{sec:background}, this number is expected to be relatively large for fast flux domains. 
\item \textit{TTL} This feature is defined as the TTL value for the DNS reply message for a given domain. As explained in Section \ref{sec:background}, fast flux domains usually have very short TTL values. 
\end{enumerate}

Table \ref{features} summarizes the main features that are used in the proposed system for fast flux detection.
In total, $8$ features are used. The system obtains the features from the DNS response message or from two databases that are stored locally. The information in the databases are obtained from Censys and IP geolocation services. Clearly, it is possible to perform the online and highly accurate fast flux detection using the proposed approach . 

\begin{table}
	\caption{The main features that are used in the system}
	\label{features}       
	\begin{tabular}{lll}
		\hline\noalign{\smallskip}
		first & second & third  \\
		\noalign{\smallskip}\hline\noalign{\smallskip}
		F1      & DomainLength             &   The length of a domain name          \\
		F2      & Regions            &  The number of countries where the IPs are located           \\
		F3      & Ports             &   The overall number of open ports for all of the IPs         \\
		F4      & IPCount             & The number of IP addresses in a DNS response message        \\
		F5      & IP ratio            &   The ratio of the returned IPs (by Censys) to the number of IPs in the DNS response message            \\
		F6      & TTL             &   The TTL value of the DNS response message   \\
		F7      & ASN             &    The ratio of the number of distinct ASNs to the number of IPs in the DNS response message    \\
		F8      & RegionalSpread             &   The ratio of the number of distinct countries to the number of IPs the DNS response message       \\
		\noalign{\smallskip}\hline
	\end{tabular}
\end{table}

\section{Classification Methods \label{sec:classification}}
Artificial Neural Network (ANN) models have been used widely for classification of data into classes. These models are powerful tools for supervised machine learning where a model can classify a data input into one of classes. During the training of an ANN model, data inputs are vectors of values and their corresponding classes. In our learning approach, vectors of the selected features along with their classes (Fast Flux or Legitimate domain) have been used to train the ANN models. 10-fold method is used on the input data for training and validation. Three types of ANN models were used which are: Multilayer Perceptron (MLP) or Feedforward Neural Network, Radial Basis Function Network (RBF), and Support Vector Machines (SVM). SVM outperformed the other two methods. Below is a brief review about each one of the ANNs. 

\subsection{MLP and RBF neural networks}
\begin{figure}
		\centering
		\includegraphics[width=\columnwidth]{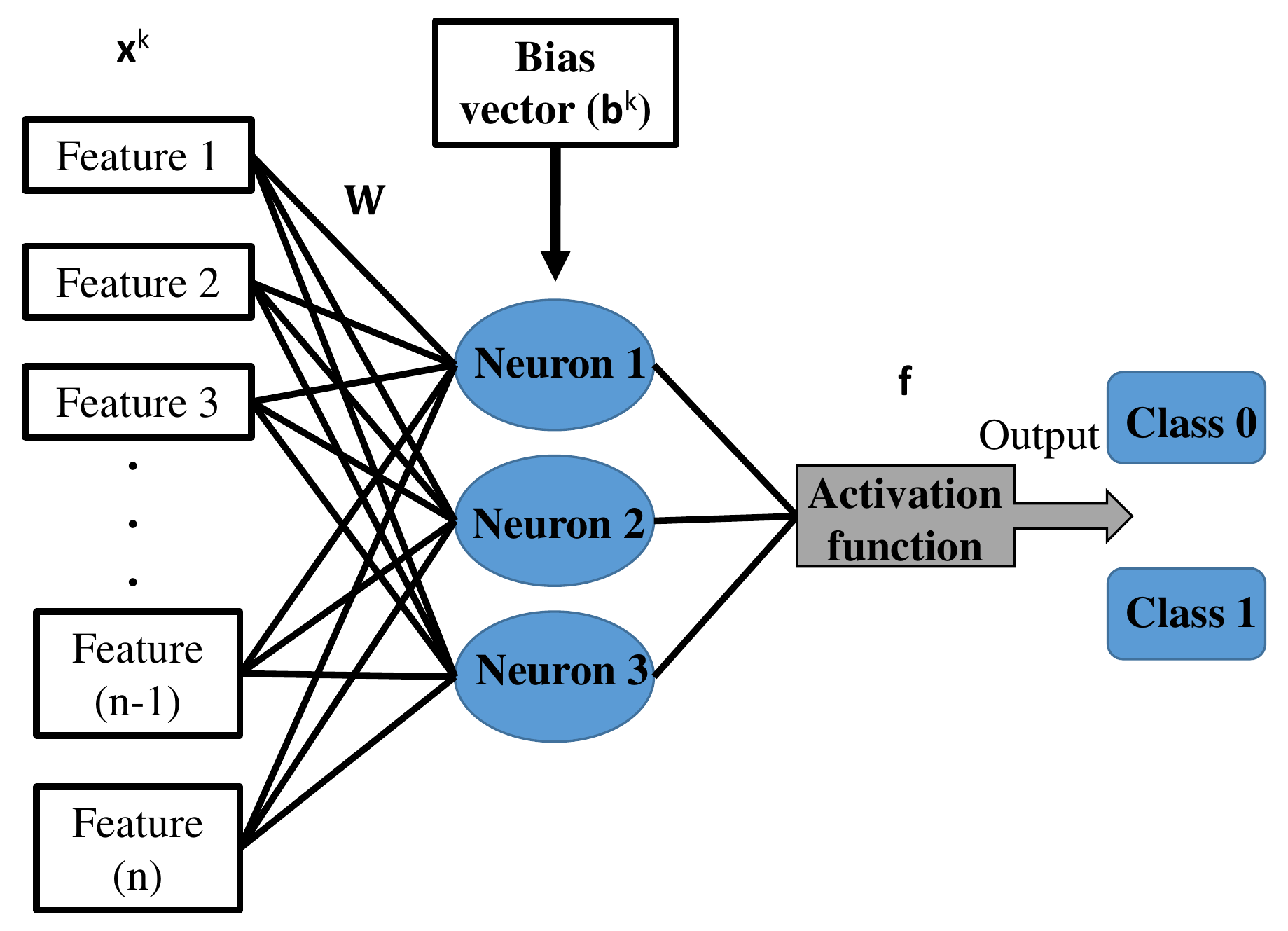}
		\caption{ Structure of a Multilayer Perceptron Network}
		\label{MLP_arch}
	\end{figure}

In MLP, the feature input vector $k$ ($\textbf{x}^k$) is multiplied by the weight matrix ($\textbf{W}$) and a bias vector ($\textbf{b}^k$) is added to the product \cite{Russell2009}. Then an activation function $\textbf{f}$ is applied to rescale the result to a value between 0 and 1. Different activation functions can be used such as the Sigmoid and the Softmax functions \cite{MLP1} \cite{MLP2}. Figure \ref{MLP_arch} depicts a general MLP network. Hence, the output of the network is calculated using Equation \ref{MPL:output}.
\begin{equation}\label{MPL:output}
    output = \textbf{f} (\textbf{W}\textbf{x}^k + \textbf{b}^k)
 \end{equation}

Softmax function has also shown high performance when it is used as the activation function. Softmax function is given in Equation \ref{softmaxf} and has been used for the output layer in our experiments.

\begin{equation}\label{softmaxf}
    g(a_x) = \frac{exp(a_x)}{\displaystyle\sum_{i=1}^K exp(a_i)} 
\end{equation}

  On another hand, Radial Basis Function (RBF) networks applies radial functions $\textbf{h}(.)$ on the input vector ($\textbf{x}^k$). A popular radial function that is widely used is the Gaussian function which is shown in Equation \ref{Gaussf}. $c$ is a center point where the function decreases or increases monotonically based on the distance from $c$. $r$ is the function radius.  

  \begin{equation}\label{Gaussf}
    \textbf{h}(x) = exp\left( - \frac{(x-c)^2}{r^2} \right)
  \end{equation}
  
\begin{figure}
		\centering
		\includegraphics[width=\columnwidth]{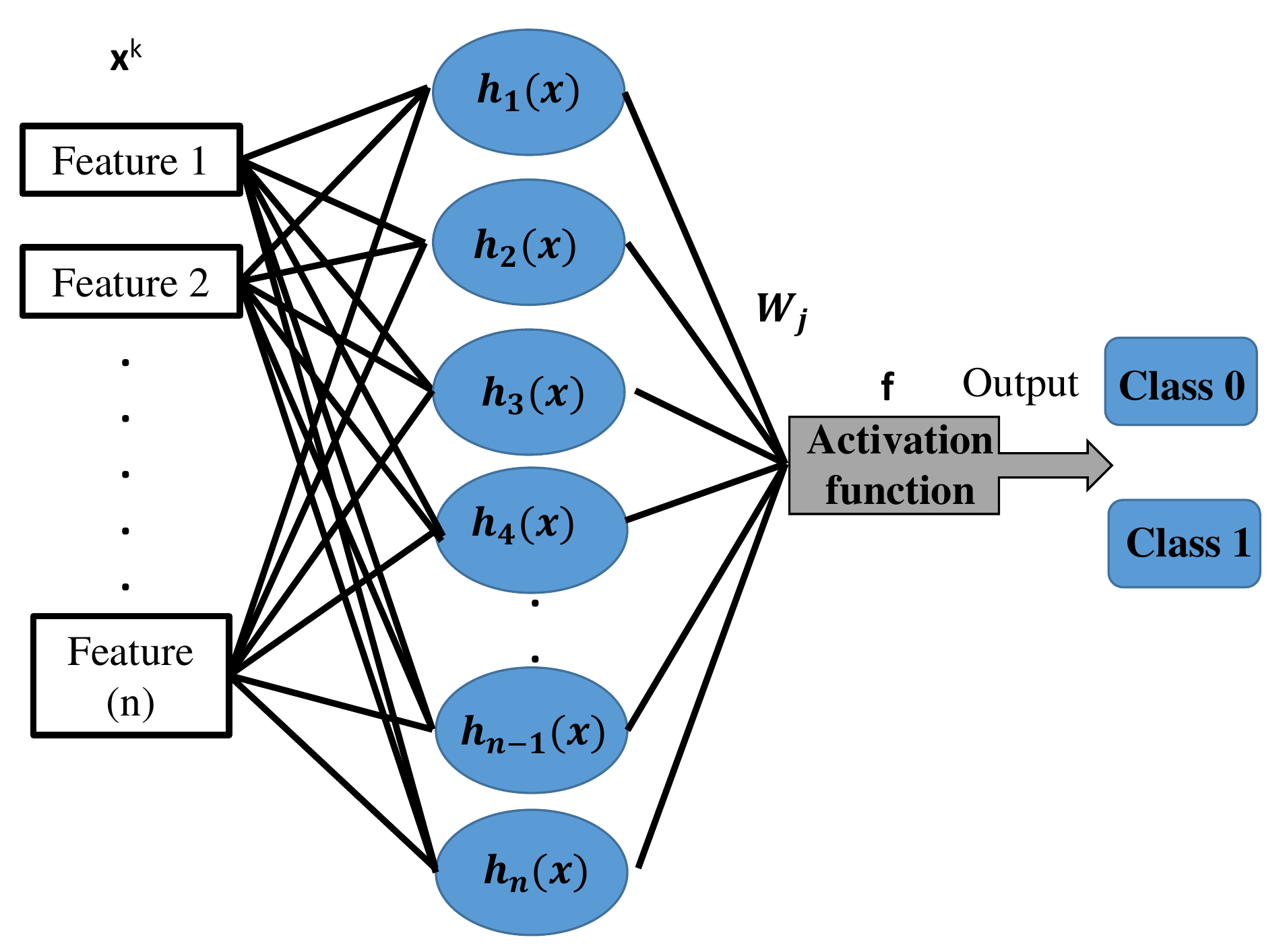}
		\caption{Structure of a Radial Basis Function Network}
		\label{RBF_arch}
	\end{figure}

  In RBF networks, the hidden layer represents the radial functions as shown in Figure \ref{RBF_arch}. Softmax or Gaussian activation functions can be used. Both functions were used in our experiments as we will discuss in the evaluation section. If Gaussian basis function is used, the output of the network is computed using Equation \ref{RBF:output} \cite{RBFNN1} \cite{RBFNN2} \cite{RBFNN3}. 

\begin{equation}\label{RBF:output}
    output = \displaystyle\sum_{j} \left( \textbf{w}_j \times exp\left( - \frac{(x-c)^2}{r^2} \right) \right)
  \end{equation}
  IBM SPSS tool was used in our experiments for both MLP and RBF nural networks. The identity function $\textbf{g}(x) = x$ is used for the output layer of the RBF network. 

\subsection{SVM and RBF kernel}

Support Vector Machine (SVM) finds a separating hyperplane with the maximum margin of separation. In other words, it computes the maximum distance between the separating hyperplane and the closest data points (support vectors). The design of a single neuron for fast flux detection can be interpreted as a classification problem. Therefore, the synthesis of fast flux detection neural network can be solved by a set of $n$ independent SVMs; where $n$ is the number of neurons in the neural network \cite{Burges1998}.

During training of the SVM ANN, ($\textbf{x}^{k}$, $y_{k}$), $k=1, 2, …, m$ represent the feature training patterns. Each feature input vector $\textbf{x}^k \in \mathbb{R}^n$ belongs to one of two classes ($y_{k}$ = -1 or +1). -1 is a fast flux domain and +1 is a legitimate domain. Assuming the feature vectors are linearly separable, there exists a separating hyperplane as depicted in Equation \ref{SVM:eq1}.

\begin{equation}\label{SVM:eq1}
     y_k (\textbf{w}^T\textbf{x}^k + \textbf{b}) > 0,   k = 1, 2,... , m
  \end{equation}
  
  The weights $\textbf{w}$ and the bias $\textbf{b}$ vector can be rescaled to get Equation \ref{SVM:eq2}. 
  
\begin{equation}\label{SVM:eq2}
     y_k (\textbf{w}^T\textbf{x}^k + \textbf{b}) \geq 0,   k = 1, 2,... , m
  \end{equation}
  
  The corresponding weights and bias represent the optimal hyperplane. To compute the optimality using Lagrange multipliers $\alpha_1 ... \alpha_m$, the objective function is to minimize Equation \ref{SVM:eq3}.
  
 \begin{equation}\label{SVM:eq3}
\textbf{J}(\textbf{$\alpha$}) = \displaystyle\sum_{k=1}^m \alpha_k - \frac{1}{2} \displaystyle\sum_{k=1}^m \displaystyle\sum_{j=1}^m \alpha_k \alpha_j y_k y_j (\textbf{x}^k)^T \textbf{x}^j~~~~~~~
\end{equation}
 
  subject to:
  \begin{equation}\label{SVM:eq4}
  \begin{array}{l}
  \displaystyle\sum_{k=1}^m \alpha_k y_k = 0 \\
  ~ \\
  \alpha_k \geq 0, k = 1, 2, ...,m
  \end{array}
  \end{equation}
  
  Solving Equation with its constraints, the optimum Lagrange multipliers are used to compute the matrix w for all neurons as depicted in Equation \ref{SVM:eq5}. For more details, please refer to \cite{ Haykin1998} and \cite{Casali2006}.
  \begin{equation}\label{SVM:eq5}
        \textbf{w} = \displaystyle\sum_{k=1}^m \alpha_k y_k \textbf{x}^k 
  \end{equation}

However, after performing our experiments, the feature input vectors are not linearly separable. To solve this problem, different kernel techniques can be used along with the SVM to map the input vectors into higher dimensions using a non-linear mapping function in order to make them linearly separable. The result is a feature space $\phi(\textbf{x})$ of the input vectors. Introducing the feature space function into Equation \ref{SVM:eq3}, the objective function can be rewritten as in Equation \ref{SVM:eq6}. 
\begin{equation}\label{SVM:eq6}
    \textbf{J}(\textbf{$\alpha$}) = \displaystyle\sum_{k=1}^m \alpha_k - \frac{1}{2} \displaystyle\sum_{k=1}^m \displaystyle\sum_{j=1}^m \alpha_k \alpha_j y_k y_j \phi(\textbf{x}^k)^T \phi(\textbf{x}^j)~~~~~
  \end{equation}
  
  A kernel function notation $\textbf{k}(\textbf{x}^k,\textbf{x}^j)= \phi(\textbf{x}^k)^T \phi(\textbf{x}^j)$ can be used in Equation \ref{SVM:eq6}, which results in Equation \ref{SVM:eq7} \cite{sharma2018} \cite{Cho2008}. 
  \begin{equation}\label{SVM:eq7}
    \textbf{J}(\textbf{$\alpha$}) = \displaystyle\sum_{k=1}^m \alpha_k - \frac{1}{2} \displaystyle\sum_{k=1}^m \displaystyle\sum_{j=1}^m \alpha_k \alpha_j y_k y_j \textbf{k}(\textbf{x}^k,\textbf{x}^j)~~~~~
  \end{equation}
  
  We have used different kernel functions (feature mappings) and the best results were achieved by the Radial Basis Function (RBF) kernel. The RBF kernel is depicted in Equation \ref{SVM:eq8}. $\textbf{SVM}^{light}$ tool was used in our experiments which provides different kernels \cite{Joachims99}. 
  
  \begin{equation}\label{SVM:eq8}
        \textbf{k}(\textbf{x}^k,\textbf{x}^j)= exp(-\gamma \left\|\textbf{x}^k-\textbf{x}^j)\right\|^2)
  \end{equation}

\section{Evaluation \label{sec:evaluation}}

IBM SPSS tool was used for both MLP and RBF neural networks under Windows 10 platform machine. For SVM, we have used $\textbf{SVM}^{light}$ tool under Ubuntu 18.04 LTS machine. The tool provides different kernel implementations \cite{Joachims99}. The dataset for conducting the experiments is described in subsection \ref{dataset}. Subsection \ref{performance} discusses the performance of different ANN techniques.  

\subsection{The dataset \label{dataset}}

For the evaluation of PASSVM, and taking into consideration that our system relies on information provided by Censys search engine about IP addresses that correspond to fast flux domains, and because Censys started to provide access to their daily IPv4 full address scans since late 2017, we restricted our fast flux dataset to includ only fast flux domains that have appeared during 2018 and 2019. In this work, we mainly collected fast flux domains that were active during April 2018 to January 2019. To achieve this goal, a seed of 80 confirmed fast flux domains were collected manually from recently published papers \cite{AGD} \cite{fastflux-passive-2}. In addition, some domains appeared in recent tweets about new fast flux domains. Then, we performed active DNS lookup for every unique domain name in the initial list using the Linux dig utility for a period of two months. Based on the DNS response messages of the DNS lookups, the list of the IP addresses of the domain names were extracted. For each IP address that was resolved from the initial fast flux dataset, a query to VirusTotal \cite{VirusTotal} was performed in order to get an updated list of the domain names that had been resolved since April 2018 along with the date associated with each domain name. After that, for each domain name, a query is sent to VirusTotal get the list of the IP addresses that were resolved for the the associated domain names and their dates. In total, we were able to obtain $5062$ fast flux domains. The dataset of the legitimate domains were obtained by performing active DNS query for Alexa's top $1$ million domains \cite{Alexa}. Then, we filtered the domain names and included only domains with $5$ or more IP addresses in their DNS response messages. The result is a dataset of $3087$ legitimate domain names and their corresponding IP addresses.   

\begin{figure}
	\centering
	\includegraphics[width=\columnwidth]{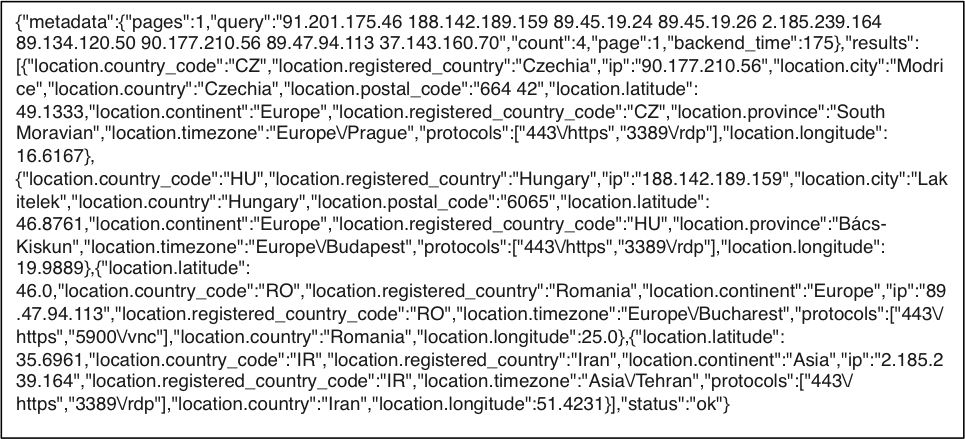}
	\caption{A sample of a JSON object returned by Censys}
	\label{jason}
\end{figure}

For each domain and its corresponding IP addresses in the aforementioned datasets, a query is submitted to Censys search engine using APIs. The results are received as JSON objects similar to the example shown in Figure \ref{jason}. These objects were parsed out to get the information of interest such as the number of distinct port numbers, the number of IP addresses, the number of countries, cities, etc. Moreover,  we used the geolocation database available at \cite{iptoasn} to get the ASN numbers and the countries that are associated with all of the IP addresses of a given domain name.

\subsection{Performance Evaluation \label{performance}} 

\begin{figure}
		\centering
		\includegraphics[width=\columnwidth]{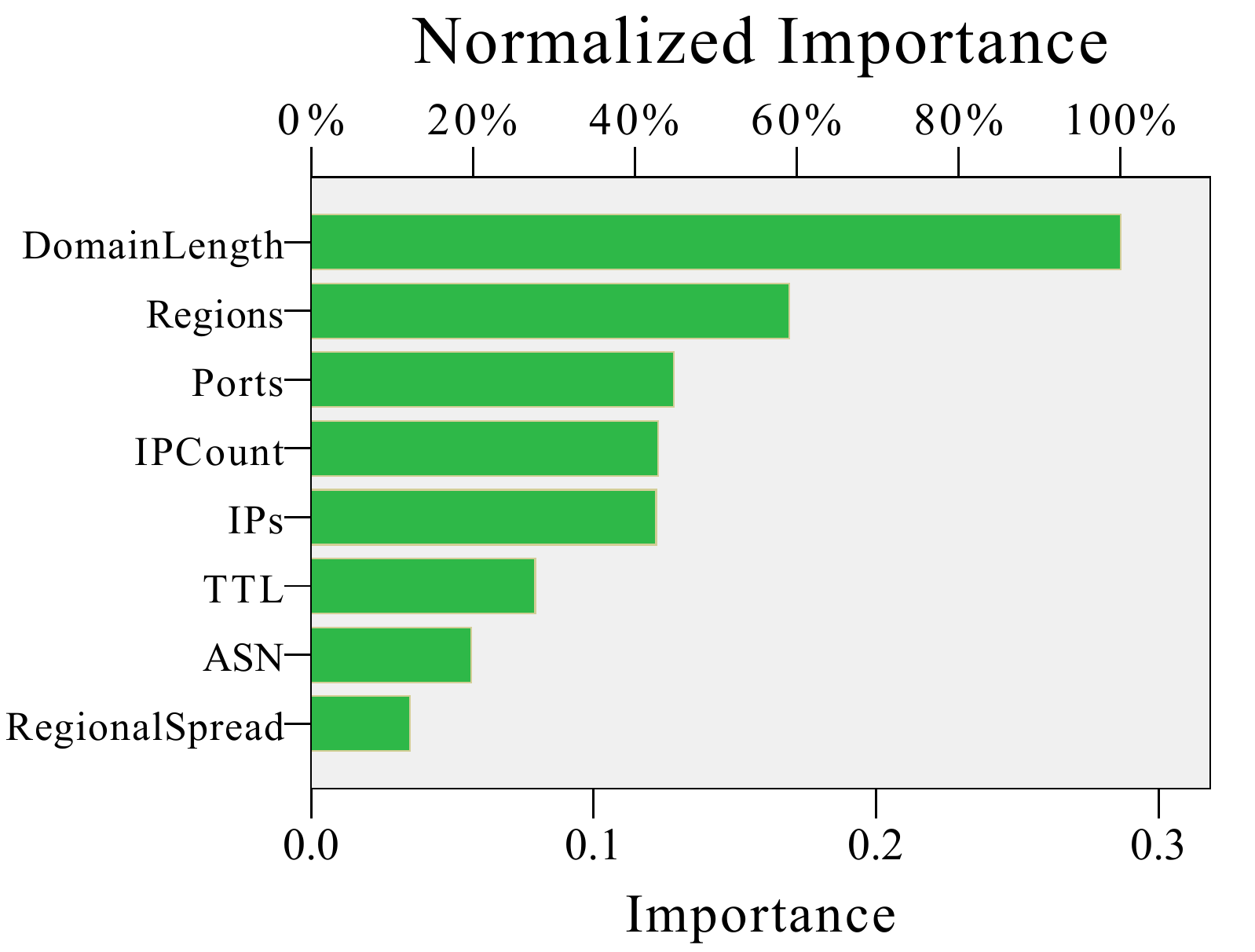}
		\caption{Normalized weight of features in MLP Network}
		\label{Comparison-MPL}
	\end{figure}

\begin{figure}
		\centering
		\includegraphics[width=\columnwidth]{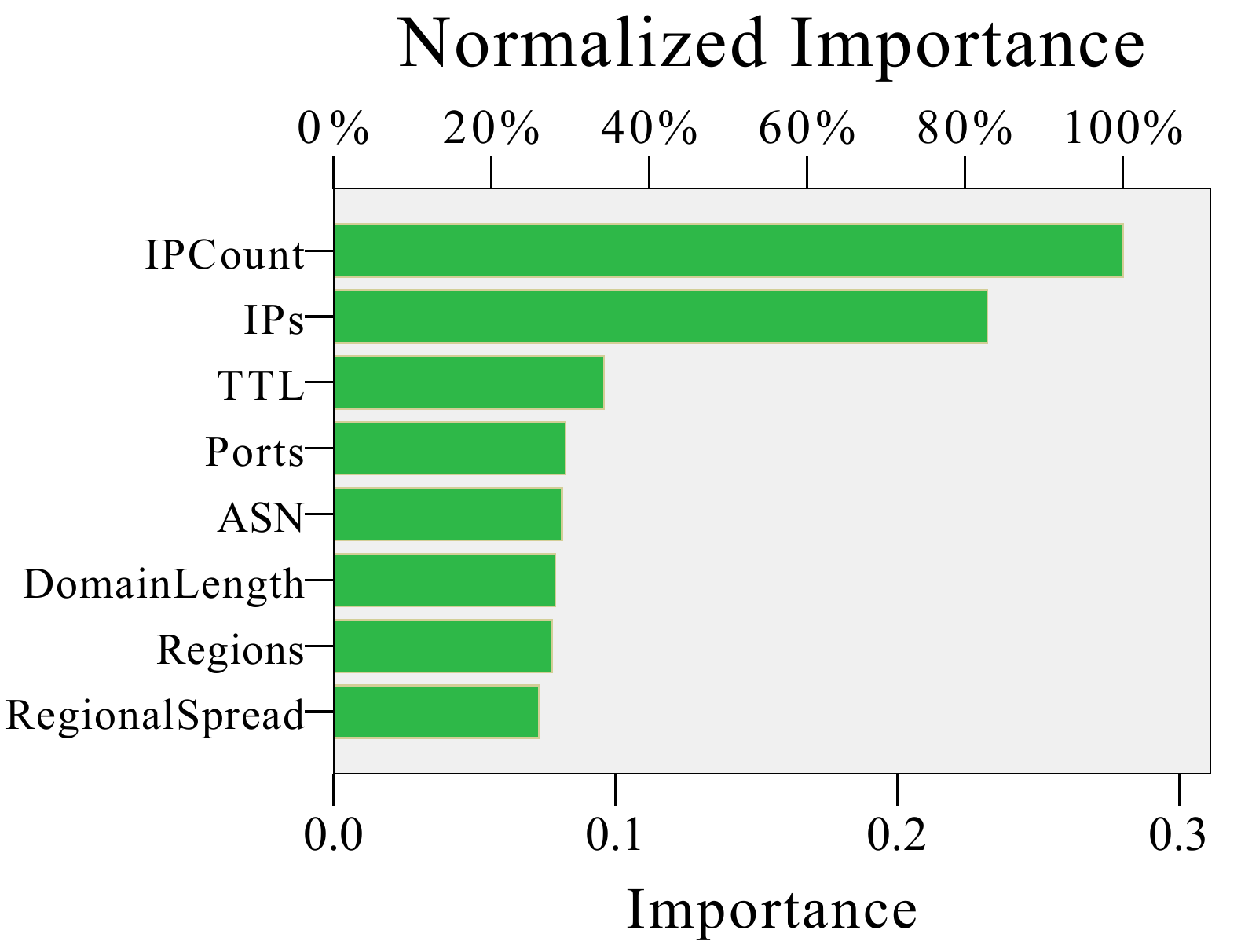}
		\caption{Normalized weight of features in RBF Network}
		\label{Comparison-RBF}
	\end{figure}

The proposed PASSVM system has been evaluated using Multilayer Perceptron (MLP), Radial Basis Function Network (RBF), and Support  Vector Machines (SVM) as described in Section \ref{sec:classification}. The feature selection is considered as an important step in training artificial neural networks. Therefore, the normalized importance of the different fast flux network features is considered in our study and was evaluated during the training experiments. Hence, the normalized importance of features for MLP network and RBF network are shown in Figure \ref{Comparison-MPL} and Figure \ref{Comparison-RBF}, respectively. Based on Figure \ref{Comparison-MPL}, it can be noted that the DomainLength, Regions, and Ports have the highest importance. However, for RBF network, the features IPCount, IPs, TTL and Ports have the highest importance as shown in Figure \ref{Comparison-RBF}. This strongly suggests that all the selected features that we have chosen in this study are important and play a major role in distinguishing fast-flux domains from legitimate domains. It also provides an evidence that the two newly introduced features (the number of open ports and the IP ratio) are of high importance to achieve highly accurate domain name classification.

\begin{figure}
		\centering
		\includegraphics[width=\columnwidth]{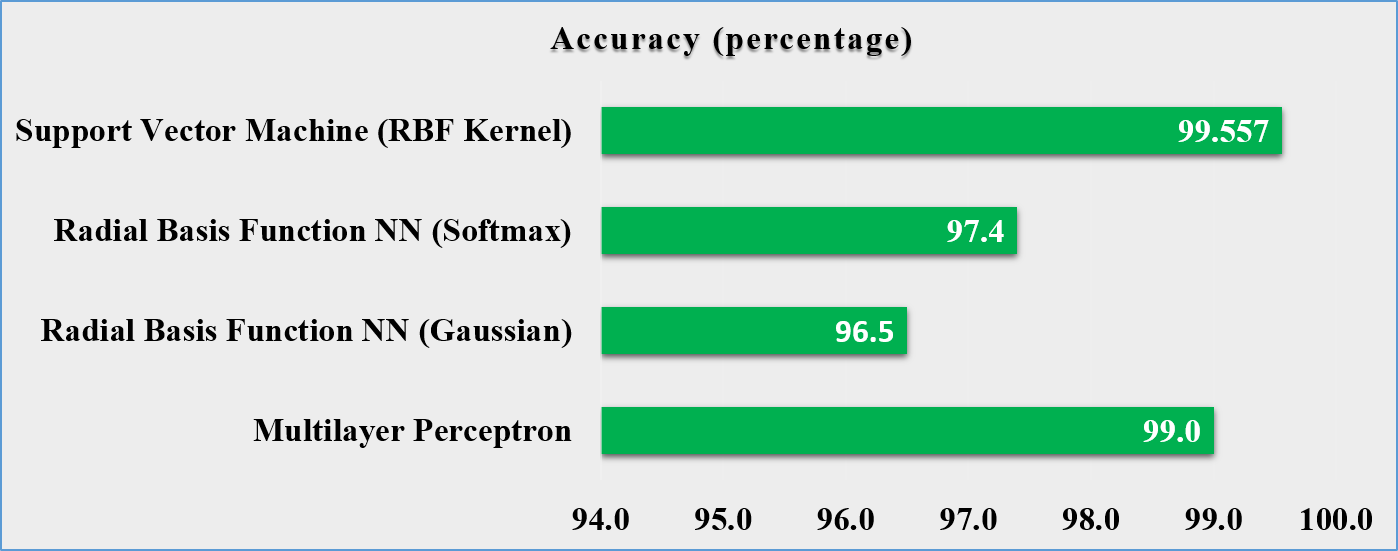}
		\caption{The accuracy of different ANN techniques}
		\label{accuracy}
	\end{figure}

We have performed several experiments to evaluate the performance of the different ANNs using the 10-fold cross-validation method in order to obtain the training model. In every run of the experiment, about 90\% of the dataset is trained, and 10\% of the dataset is tested. Results show that the Support Vector Machine with RBF kernel outperforms other ANNs with an accuracy of $99.557$ as shown in Figure \ref{accuracy}. Table \ref{accuracy-FPR} compares the accuracy, false positive rates, and false negative rates of the different classifiers. It can be seen that SVM with RBF kernel has the highest performance in terms of accuracy and false negative rate with a very low false positive rate. Such a very low false negative can be cached in a lookup table for a practical adaptation PASSVM. It is important to point out that the overall average execution time of this classifier for training and testing does not exceed 40 ms, and the average time to test one DNS record was less than 18 ms. After the comparison with other ANN algorithms, SVM with RBF kernel is chosen because of its practicality in performing very fast and highly accurate fast flux detection.

\begin{table}
	\caption{Comparison of different ANN classifiers}
	\label{accuracy-FPR}       
	\begin{tabular}{llll}
		\hline\noalign{\smallskip}
		Classifier       & Accuracy & FPR  & FNR   \\
		\noalign{\smallskip}\hline\noalign{\smallskip}
		SVM (RBF Kernel) & 99.557   & 0.008 & 0.004  \\ 
		
		RBF (Softmax)    & 97.4     & 0.0058  & 0.1001  \\ 
		
		RBF (Guassian)~  & 96.5     & 0.0066   & 0.1284  \\ 
		
		MLP              & 99.0     & .0048   & 0.0483    \\
		\noalign{\smallskip}\hline
	\end{tabular}
\end{table}

\subsection{Comparison with the state-of-the-art}
Table \ref{comparison} presents a comparison between the proposed PASSVM system, which is based on SVM with RBF kernel, with the state-of-the-art mechanisms for fast flux detection. The mechanisms include GRADE \cite{grade}, FF-Hunter \cite{fastflux-passive-5}, FluxBuster \cite{FluxBuster} and \cite{AGD}. The comparison criteria are based on whether the detection is performed online or offline, the capability of performing a detection based on a single DNS record, the time for training and testing, the accuracy, and the used memory that were reported by the authors of the methods. 

As shown in Table \ref{comparison}, PASSVM has the best performance in terms of time, accuracy, and used memory. Also, it has the capability of performing online fast flux detection based on a single DNS packet given that the system is trained in advance. This allows for the detection of new fast flux domains as they appear in the wild when users tries to access them. Hence, PASSVM can be used by organizations to provide a passive online fast flux detection. In addition, the system AI-model can be trained again and again as more dataset becomes available to enhance its accuracy. 

Previously proposed systems take more time to classify a domain name because they require features that should be obtained from different Internet sources rather than from local databases. For example, GRADE performs the fast flux detection task based on the entropy of domains proceeding nodes for all A records, and the standard deviation of the round trip time for all of the A records. This requires performing tracerout and real-time measurement of the round trip time for all of the records, which incurs high overhead and has a major problem of possible failure due to filtering of the ICMP messages. The system proposed in \cite{AGD} analyzes live traffic that is collected from the upper DNS hierarchy by applying literal composition to identify DGA-generated domains. Then it  clusters the domains based on their literal features and the edit-distance. Extreme machine learning (EML) is used to classify the domain clusters into fast-flux domains and legitimate domains based on different features that require active query of the whois database. Some of the 14 features that were used in FF-Hunter require collecting a large number of DNS records for a given domain, which means that it can not perform detection based on a single DNS record as it takes extra time to collect the required features. FluxBuster \cite{FluxBuster} relies on characteristics that are obtained from passive DNS traffic traces, in addition to active data collection in order to accurately perform domain names classification. Moreover, it does not provide fast detection and requires the collection of a large number of DNS records for each domain.

\begin{table}
	\caption{Comparison with the state-of-the-art systems}
	\label{comparison}       
	\begin{tabular}{llllll}
		\hline\noalign{\smallskip}
		Algorithm    & Online? & Single Packet Detection? & Time used (s) & Accuracy & Used memory (MB)  \\
		\noalign{\smallskip}\hline\noalign{\smallskip}
		GRADE \cite{grade}        & yes     & yes                      & 30.48        &   98.47       & 85.34             \\
		FF-Hunter \cite{fastflux-passive-5}    & yes     & NO                      & 43.87        &    98      & 92.71             \\
		FluxBuster \cite{FluxBuster}  & No      & NO                      & 198.45       &   99.4       & 102.32            \\
		Zang, Gong, Mo, Jakalan, Ding \cite{AGD}         & yes     & NO                      & 32.71        &   99.1       & 105.64            \\
		PASSVM & yes     & yes                     & 0.04         & 99.557   &  40.00                 \\
		\noalign{\smallskip}\hline
	\end{tabular}
\end{table}

\section{Conclusions \label{sec:conclusion}}
Fast flux service networks provide Internet adversaries with the capability to hide their malicious servers while maintaining a high availability. There is a pressing need to identify fast flux networks in a short time in order to minimize the risk of accessing malicious websites and hence spreading malware. In this paper, a novel AI-based online fast flux detection system is proposed. The proposed PASSVM system applies artificial intelligence algorithms to identify fast flux domains based on features that are associated with a single DNS record. The features are obtained directly from the record itself, in addition to information that is available in local databases. The databases information are obtained from the Censys search engine and IP geolocation service. PASSVM system performs online fast flux detection with high accuracy. Experimental evaluations demonstrate that SVM with RBF kernel outperforms other artificial neural networks and achieves high accuracy of 99.557\% with a low false negative rate of 0.4\%. Such a low rate can be cached in a lookup table for practical adaptation of the system in order to achieve a zero rate. Compared with the state-of-the-art fast flux detection systems, PASSVM achieves the best performance in terms of accuracy, time, and used memory. Also, the system approach makes it practical to be employed within organizational networks so that employees do not access malicious domains, and hence it prevents the spread of malware infections.

\section*{Acknowledgement}
We thank Censys team for providing us with an access to the Internet wide scanning database in order for us to conduct the research. We also thank VirusTotal team for providing us a private API key to access their data for collecting the fast flux dataset.

\appendix


\bibliographystyle{cas-model2-names}


\textbf{Basheer Al-Duwairi} is an Associate Professor at the Department of Network Engineering and Security at Jordan University of Science \& Technology. He received his B.S. in electrical and computer engineering from Jordan University of Science and University (JUST) in 1999, and his M.S. and PhD in computer engineering from Iowa State University, Ames, IA in 2002 and 2005, respectively. Over the past 15 years, Al-Duwairi's research interests are in the area of network security focusing mainly on developing efficient schemes for DDoS mitigation, Botnets, Email spam filtering, and studying the emerging threats in new network architectures. \\\\
\textbf{Moath Jarrah} is an Associate Professor at the Department of Computer Engineering at Jordan University of Science and Technology (JUST). He has received his Master and Ph.D. degrees from the Department of Electrical and Computer Engineering at the University of Arizona, Tucson, USA, in 2005 and 2008, respectively. He received the B.S degree from the Department of Electrical and Computer Engineering at JUST in 2002. Jarrah worked as a senior research fellow at Rolls-Royce corporate lab at Nanyang Technological University (NTU), Singapore; and as a scholar researcher at the Department of Creative IT Engineering and at POSTECH University, South Korea in the period between 2014-2016. He has co-authored and published many papers in top-tier journals and conferences. His research interests include modeling and simulation, machine learning, cloud computing, smart grid, and optimization. He is a reviewer of many international conferences and journals.\\\\
\textbf{Ahmed Shatnawi} is an Assistant Professor in the Software Engineering department at Jordan University of Science and technology. Ahmed’s research interests lie primarily in the intersection of software engineering and information security. His research focuses on finding better ways to design software systems that are safe, secure, and reliable to use. He received his Ph.D. in Engineering from the University of Wisconsin Milwaukee in 2017 and his M.S. in Software Engineering from George Mason University in 2012.

\end{document}